\documentclass[twocolumn,showpacs,preprintnumbers,amsmath,amssymb]{revtex4}
\usepackage{graphicx}
\usepackage{dcolumn}
\usepackage{amsmath,amsfonts}
\usepackage{latexsym} 
\usepackage[dvips,final]{epsfig}

\begin{document}

\title{Optimum control of broadband noise by barriers based on sonic crystals}
\author{Victor M. Garcia-Chocano}
\author{Jos\'e S\'anchez-Dehesa}
\email[]{Corresponding author: jsdehesa@upvnet.upv.es}
\affiliation{Wave Phenomena Group, Departamento de Ingenieria Electr\'onica, Universidad Polit\'ecnica de Valencia, Camino de vera s.n., E-46022, Spain
}
\date{\today}

\begin{abstract}
It is demonstrated that sonic crystals (periodic structures of sound scatterers) can be used to design acoustic barriers that attenuate efficiently broadband noise. Traffic noise is chosen here as an example in which our design procedure is applied. The structures consist of cylindrical units containing rubber crumb, a sound absorbing material. An optimization algorithm is developed to obtain the material distribution and the dimensions of the sonic crystal giving the best attenuation properties for this noise. The good agreement found between predictions and measurements for a barrier (3m height) characterized in a transmission room gives strong support to our proposal.
\end{abstract}


\maketitle
\section{\label{sec:intro}Introduction}
Attenuation of broadband noise is a topic of increasing interest because of the damage producing on human behavior and social environment. 
Broadband noise is generated by industrial machinery and many other products that we use in our modern lifestyle. Particularly, traffic noise is perhaps a paradigmatic example of broadband noise where different approaches for its attenuation have been extensively researched; see, for example the reviews in \onlinecite{Kur74,Eki03} and references therein. 
Since the late nineties, two-dimensional sonic crystals were proposed \cite{Sig96,San98,Che01,San02,Umn05} as an alternative to conventional noise barriers. 
Sonic crystals consist of periodic arrangements of scatterers that inhibit sound transmission for certain ranges of frequencies called bandgaps, as photonic crystals do with light. 
Optimization algorithms have been applied to this type of structures in order to enhance their acoustic performance. 
For example, H\aa kansson and coworkers applied genetic algorithms to design flat acoustic lenses\cite{Hak05}, demultiplexers \cite{Hak06} and many others acoustic devices. 
The same optimization procedure was recently used to enhance the attenuation properties by means of the creation of vacancies in the periodic structures \cite{Rom09}. 
However optimization procedures recently studied in noise barriers design based on periodic structures have two main drawbacks. First, the barriers are very thick structures 
(i.e., consist of many layers of scatterers)  making them practically unfeasible for their construction due to their cost and space requirements.
Second, the noise barriers that are placed at opposite sides of the traffic road are not taken into account. Under this circumstance multiply reflected sound between parallel barriers can cause a significant increase in noise in the screened area \cite{Wat96}. 

This work introduces a design procedure of sonic crystal-based barriers intended to attenuate efficiently broadband noise. The procedure is specifically applicable to traffic noise, which is a class of broadband noise that has been very well characterized and widely studied. 
The barriers are based on the combination of two attenuation mechanisms; one is sound attenuation at the Bragg frequencies (due to reflectance) and the second is sound absorption by the material employed in the sonic crystal building-units. The barriers here introduced are made of only three rows of cylindrical units that contain rubber crumb, a sound absorbing material obtained from recycling car tires. An optimization algorithm is employed to obtain the optimum dimensions and material distribution in the scattering units of each row, as well as the distances between scatterers. A selected barrier has been constructed and experimentally characterized to support our proposal. 
Let us point out that the approach here applied to attenuate broadband noise generated by traffic noise can be easily extended to other types of broadband noise provided that its spectral profile is previously characterized.

The paper is organized as follows. Section \ref{sec:model} briefly describes the multiple reflection model applicable to the double-barrier configuration. 
Section \ref{sec:OpModel} presents the optimization goal function and the optimization algorithm. 
Results predicted by the optimized design are discussed in Sec. \ref{sec:RD}. 
The experimental validation of the algorithm is reported in Sec. \ref{sec:Experiment}, where we present a physical realization of a barrier of 3 meter height. 
Finally, Sect. \ref{sec:Conclusion} summarizes the work and gives conclusions. 

%
\section{\label{sec:model}Multiple reflections model}
\begin{figure}
\includegraphics[width=8.5cm]{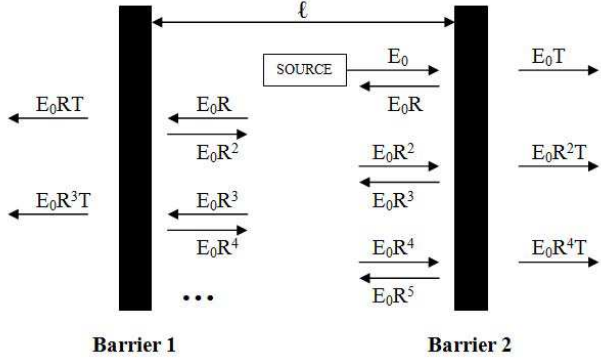}
\caption{Scheme of the multiple reflection model. The power in $R$ defines the number of reflections experienced by the sound wave with initial energy $E_0$.}
\label{fig:MultiplesRebotes}
\end{figure}

Consider two barriers placed at both sides of a road where the traffic noise is generated. 
By assuming an incoherent source, the problem of multiple reflections in the double barrier configuration can be approached with a ray model where the total energy at each point corresponds to the sum of the energy of every arriving path \cite{Hur80}. 
The simplified one-dimensional model is described in Fig. \ref{fig:MultiplesRebotes}, 
where the noise source (the car) radiates a plane wave with energy $E_0$. 
Each time the wave impinges a barrier, a fraction $T$ of the incident energy is transmitted through the barrier, another fraction $R$ is reflected and the rest $1 - R - T$ is absorbed. The effective transmission coefficient, $T_{eff}$, is defined as the fraction of energy transmitted through both barriers:
\begin{eqnarray}
\label{Teff}
T_{eff}=T\sum_{n=0}^{+\infty}{R^n}=\frac{T}{1-R}
\end{eqnarray}
This simplified model can be enhanced by considering the case of an infinitely long and straight road where continuous traffic approximately behaves as a line source in such a manner that, at enough distance from the individual sources, sound levels fall 3dB per doubling of distance as cylindrical wavefront does \cite{Emb96}. 
In other words, an attenuation proportional to $1/r$ (being $r$ the traveled distance) is finally obtained due to geometrical spreading. 
The corresponding transmission coefficient is
\begin{eqnarray}
\label{Teff2}
T_{eff}^\prime=2\frac{T}{\ell}\sum_{n=0}^{+\infty}{\frac{R^n}{2n+1}}=2\frac{T}{\ell}\frac{\operatorname{artanh}\left(\sqrt{R}\right)}{\sqrt{R}},
\end{eqnarray}
where $\ell$ is the separation between barriers. 

Another model is to consider road vehicles acting like individual point sources with spherical spreading; i.e. the sound energy is decreased by a factor $1/r^2$ the coefficient transmission becomes:
\begin{eqnarray}
\label{Teff3}
T_{eff}^{\prime\prime}=4\frac{T}{\ell^2}\sum_{n=0}^{+\infty}{\frac{R^n}{(2n+1)^2}}.
\end{eqnarray}

In fact there are numerous traffic noise models, some of them considering the noise source as incoherent punctual sources and other as a line source \cite{Ste01}. 
Because of this, the three canonical cases described by Eqs. [\ref{Teff}]-[\ref{Teff3}] are analyzed here.

%
\section{\label{sec:OpModel}Optimization algorithm}
We chose to maximize the so called insulation index for airborne sound ($DL_R$), which is used in Europe as a criterion to classify the acoustical performance of traffic noise barriers. 
It is defined in the European Normative EN 1793-2 \cite{EN1793} as
\begin{eqnarray}
\label{DLr}
DL_R=-10\log_{10}{\left|\frac{\sum_{i=1}^{18}{10^{-0.1R_i}10^{0.1L_i}}}{\sum_{i=1}^{18}{10^{0.1L_i}}}\right|},
\end{eqnarray}
where $L_i$ is the normalized traffic spectrum, $R_i$ is the sound transmission loss and $i$ is an index indicating the eighteen standard third octave bands with frequencies from 100Hz to 5kHz. 

Parameters $L_i$ in Eq. \eqref{DLr} take into account the human hearing response and emphasizes the frequencies where the traffic noise is more undesirable, having a maximum weight into the 1kHz band.  The measurement of $R_i$ is standardized in ISO140-3 \cite{ISO140} and it is briefly explained in Sect. \ref{sec:Experiment}. 
Note that EN 1793-2 is intended to characterize the intrinsic properties of the barrier, disregarding \emph{in situ} conditions. 
Then, no discussion about diffraction by the top end of the cylinders or ground effects will be made in this work.

\begin{figure}
\includegraphics[width=5cm]{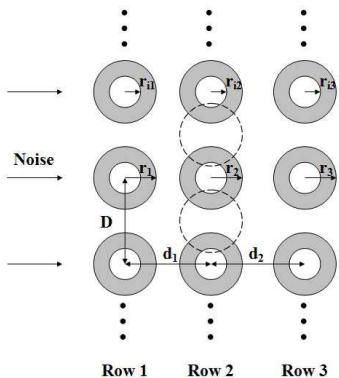}
\caption{Parameters employed in the optimization of a noise barrier based on a sonic crystal consisting of three rows of cylindrical scatterers containing rubber crumb. Shadowed regions define the location of rubber crumb. The dashed circles indicate the positions of cylinders for the opaque (hexagonal symmetry) configuration.}
\label{fig:EsquemaBarrera}
\end{figure}

%
The proposed structure to be optimized consists of a sonic crystal made of three infinite rows of cylinders, where each cylinder has an inner rigid core of radius $r_{ik}$ ($k=1, 2, 3$) and a layer of rubber crumb between the core and an external radius $r_k$ (see Fig. \ref{fig:EsquemaBarrera}). 
Each row has identical cylinders separated at a distance $D$ and the three rows are separated by distances $d_1$ and $d_2$, respectively.
 Therefore the optimization model involves 9 independent parameters. 
 Note that cylinders are aligned in a square lattice, thus lowering the flow resistance of the barrier and making it partially transparent to light. 
 For comparison purpose, we also studied a barrier in which cylinders at the middle row are displaced a distance $D/2$, being placed in the dashed circles in Fig. \ref{fig:EsquemaBarrera}. 
 Since this structure forbids the light passing through, it will be called as opaque barrier.

Calculations of barrier parameters are performed by applying the multiple scattering method. For computer performance reasons we use a 2D model, that is, infinitely long cylinders are considered. For technical details of the method the reader is referred to Ref. \onlinecite{San03} and references therein.
Within this framework it is possible to calculate the reflectance $R_{SC}$ and transmittance $T_{SC}$ of infinite sonic crystals by integrating the energy flux in a unity cell \cite{T_Tor08}, resulting: 
\begin{eqnarray}
\label{Tsc}
T_{sc}\left(\omega ,\theta_0\right)=1+2\mathfrak{R}\left [ C_0^+ \right ]+\sum_{\cos\theta_\nu \in \mathfrak{R}} \frac{\left| \cos \theta_\nu \right|}{\cos \theta_0} \left| C^+_\nu \right|^2,
\end{eqnarray}
\begin{eqnarray}
\label{Rsc}
R_{sc}\left(\omega ,\theta_0\right)= \sum_{\cos\theta_\nu \in \mathfrak{R}} \frac{\left| \cos \theta_\nu \right|}{\cos \theta_0} \left| C^-_\nu \right|^2,
\end{eqnarray}
where $\omega$ is the angular frequency and $\theta_0$ the angle of the impinging plane wave with respect to normal incidence. $C_\nu^-$ and $C_\nu^+$ are the amplitudes of the reflected and transmitted $\nu$-modes which propagate at an angle $\theta_\nu$, obtained by the expression
 \begin{eqnarray}
\label{Ov}
\sin \theta_\nu = \sin \theta_0 + \frac{2\pi\nu}{kD},
\end{eqnarray}
$k$ being the wavenumber and $\nu$ an integer number. 
Absorption by the sonic crystal barrier can be calculated by assuming energy conservation, so $A_{SC}=1-T_{SC}-R_{SC}$. 
The acoustic model used in describing the rubber crumb is based on the complex dynamical mass density and bulk modulus proposed by Johnson \cite{Joh87} and Stinson \cite{Sti92}. 
A detailed theoretical and experimental work about the application of this model within the multiple scattering method has been recently reported in Ref. [\onlinecite{San10}].

Multiple scattering method works with coherent waves with well defined wave fronts, making difficult to model the diffuse sound field required by the ISO140 norm \cite{ISO140}. 
Since this type of field is the basis of several standardized measurements, analytical predictions have been developed in control noise research. 
A conventional calculation is based on the assumption that the angular distribution of incident sound field on the surface of the sample becomes fully uniform \cite{Kut09}. Consider a surface element of the sample $dS$ and an isotropic incident sound intensity $I$. The fraction of acoustic power that arrives to $dS$ from the direction defined by the solid angle $d\Omega$ is
\begin{eqnarray}
\label{dPi}
dP_i=I\cos(\theta)dSd\Omega=IdS\cos(\theta)\sin(\theta)d\varphi d\theta,
\end{eqnarray}
where $\theta$ and $\varphi$ corresponds to the azimuth and the inclination angles in spherical coordinates, respectively. The total arriving power is calculated by integrating Eq. \ref{dPi} over the surface of a semi sphere
\begin{eqnarray}
\label{Pi}
P_i=\int_0^\pi{\int_0^{2\pi}{IdS\cos(\theta)\sin(\theta)d\varphi} d\theta}=\pi I dS,
\end{eqnarray}
Similarly, the total transmitted power is obtained as
\begin{eqnarray}
\label{Pt}
P_t&=\int_0^\pi{\int_0^{2\pi}{I\: T(\theta)dS\cos(\theta)\sin(\theta)d\varphi} d\theta} \nonumber \\
&=\pi IdS \int_0^{\pi/2}{T(\theta)\sin(2\theta)d\theta},
\end{eqnarray}
being $T(\theta)$ the angle dependent transmission coefficient, that is supposed to be independent of $\varphi$. The angle averaged transmission coefficient is given by the ratio
\begin{eqnarray}
\label{Tav_Paris}
T_{av}=\frac{P_t}{P_i}=\int_0^{\pi/2}{T(\theta)\sin(2\theta)d\theta},
\end{eqnarray}
However uniform distribution does not fully reflect the actual sound field, so an angle dependency correction based on Gaussian distribution is applied \cite{Kan00,Kan02}. 
Hence the averaged transmission coefficient of the barrier is calculated as
\begin{eqnarray}
\label{Tav}
T_{av}\left(\omega\right)=\frac{\int_{0}^{\pi/2} e^{-\theta ^2} T_{SC}\left(\omega,\theta\right) \ sin\left(2\theta\right) \, d\theta}{\int_{0}^{\pi/2} e^{-\theta ^2} \sin\left(2\theta\right) \, d\theta },
\end{eqnarray}
and the same procedure can be applied to reflection and absorption coefficients, giving $R_{av}\left(\omega\right)$ and $A_{av}\left(\omega\right)$ from $R_{SC}\left(\omega,\theta\right)$ and $A_{SC}\left(\omega,\theta\right)$, respectively.

Coefficients $R_i$ in Eq. \eqref{DLr} are calculated as $1/T_x$ where $T_x$ is the direct $T=T_{av}$ or effective $T_{eff}$, $T_{eff}^{\prime}$, $T_{eff}^{\prime\prime}$ transmission coefficients obtained by applying $T_{av}$ and $R_{av}$ to Eqs. [\ref{Teff}-\ref{Teff3}]. These parameters are calculated at several frequencies in each third octave band and then they are integrated in order to get a single $R_i$ per band.

Due to the difficulty of differentiating the 9-dimensional objective function, we have employed the Nelder-Mead optimization method \cite{Nel65} which is based on the simplex algorithm. With the purpose to globalize the search, several initializations have been programmed in order to ensure that not local maximums are reached. 
Physical constrains such us positive dimensions or no overlapped cylinders need to be taken into account in the algorithm. 
Also practical constrains like maximum barrier width of 1 meter have been imposed. 
Maximum external radii of 10 cm and distance $D$ at least four times the maximum radius are set in order to ensure a partial transparency of the barrier based on square symmetry. 
For the case of opaque barriers (hexagonal symmetry) the minimal distance between cylinders is set to 1 cm.

\begin{table}
\centering
\caption{Barrier parameters (see Fig. \ref{fig:EsquemaBarrera}) obtained from the optimization algorithm. Length dimensions are in cm. Last row contains the airborne insulation index $DL_R$. Note that the highest quality barriers, class $B_3$, according to the European normative is achieved when $DL_R > 24dB$\cite{EN1793}.}
\label{tab:ResultadosOptim}
\begin{ruledtabular}
\begin{tabular}{lcccccccc}
 &\multicolumn{2}{c}{$T$} &\multicolumn{2}{c}{$T_{eff}$} &\multicolumn{2}{c}{$T_{eff}^\prime$} &\multicolumn{2}{c}{$T_{eff}^{\prime\prime}$}\\
 &$\Box$&$\triangle$&$\Box$&$\triangle$& $\Box$ &$\triangle$& $\Box$& $\triangle$\\
$r_1$    & 10.0 & 10.0 & 10.0 & 10.0 & 10.0 & 10.0 & 10.0 & 10.0\\
$r_2$    & 10.0 & 10.0 & 10.0 & 10.0 & 10.0 & 10.0 & 10.0 & 10.0\\
$r_3$    & 10.0 & 10.0 & 10.0 & 10.0 & 10.0 & 10.0 & 10.0 & 10.0 \\
$r_{i1}$ &  4.6 & 10.0 & 3.4  &  5.1 &  4.2 &  7.3 &  4.5 &  9.2 \\
$r_{i2}$ &  4.3 &  5.0 & 4.0  &  4.3 &  4.3 &  4.1 &  4.3 &  3.8 \\
$r_{i3}$ &  4.7 &  9.5 & 4.5  & 10.0 &  4.6 & 10.0 &  4.7 &  9.4 \\
$d_1$    & 32.1 & 18.2 & 31.2 & 18.2 & 31.8 & 18.2 & 32.1 & 18.2\\
$d_2$    & 48.9 & 18.2 & 49.8 & 18.2 & 49.1 & 18.2 & 48.9 & 18.2\\
$D$      & 40.0 & 21.0 & 40.0 & 21.0 & 40.0 & 21.0 & 40.00& 21.0\\
$DL_R$(dB)& 7.2 & 18.6 &  6.7 & 16.6 & 14.0 & 24.7 & 21.1 & 32.1\\
\end{tabular}
\end{ruledtabular}
\end{table}

\section{\label{sec:RD}Results and discussion}
Barrier parameters resulting from the optimization process are shown in Table \ref{tab:ResultadosOptim}, where a distance $\ell=$10m between parallel barriers has been assumed in Eqs. \eqref{Teff2} and \eqref{Teff3}. 
It is noticed that, for the case of opaque barriers ($\triangle$ symbols in Table \ref{tab:ResultadosOptim}), the algorithm always converged to the maximum of data constrains (maximum external radius and minimum distance between cylinders); that is, trying to make the sonic crystal more compact.
The same applies to the case of semi-transparent barriers ($\Box$ symbols) where minimum distance $D$ and maximum radius are also obtained. 
In addition distances $d_1$ and $d_2$ are practically the same for each transmission model in this topology. 
\begin{figure}
\includegraphics[width=7cm]{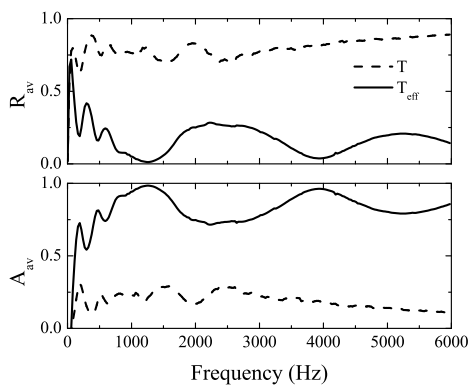}
\caption{\label{fig:ReflectAbsorc} Angular-averaged reflection $T_{av}(\omega)$ and absorption $A_{av}(\omega)$ of the opaque barriers optimized through the $T$ and $T_{eff}$ models.}
\end{figure}

Looking closer to the parameters obtained, especially the internal radii, it is observed that cases based on effective transmission $T_{eff}^{\prime}$ and $T_{eff}^{\prime\prime}$ tend to the values obtained for the case of simple transmission, $T$. 
This effect is due to the geometrical spreading of waves. 
This fact is easy to understand since multiple reflections are negligible when attenuation by wave propagation is large, becoming important only the first incidence of the wave in the barrier. 
Note that differences in $DL_R$ between the $T$ and $T_{eff}^{\prime}$, $T_{eff}^{\prime\prime}$ cases are mainly due to the attenuation in the first travel from the midpoint between barriers to the first barrier. 
Therefore, models $T$ and $T_{eff}$ can be seen as two extreme cases between which practical cases will stay depending on their attenuation by propagation and the distance between barriers. 
While in wide highways $T_{eff}^{\prime}$ and $T_{eff}^{\prime\prime}$ models will give parameters approaching those given by $T$, situations such as narrow roads or railway lines will make these parameters closer to the $T_{eff}$ model. 

It is worth to note that for the case of simple transmission $T$ there are two insulation mechanisms available: reflectance by the barrier periodic structure and absorption by the material of its building units. However the $T_{eff}$ model only employs absorption since reflected waves always reach the opposite barrier without being attenuated during propagation. The practical constrain of minimum distance $D$ imposed to the optimization of the transparent barriers has avoided the chance of using reflection as the principal attenuation mechanism (band gaps can not be formed on an efficient way). Therefore the optimization process has always chosen improving the absorption, thus resulting in barrier structures with similar parameters in the four transmission cases. On the other hand for the opaque barrier, where distances between cylinders are not so constrained, larger differences are found between the optimized parameters from the different transmission models. Note that in this configuration the barrier obtained from the $T$ model needs larger rigid cores, especially in the first row, while the barrier derived from the $T_{eff}$ model requires more rubber crumb. 
This fact highlights the differences between the transmission models in such a manner that in the first case a reflective barrier is obtained while the second corresponds to a barrier based on absorption phenomenon. 
The angular-averaged reflection $T_{av}(\omega)$ and absorption $A_{av}(\omega)$ of these two cases are represented in Fig. \ref{fig:ReflectAbsorc}, which shows how the barrier obtained from model $T$ has a high reflective spectrum but a low absorption spectrum, while that resulting from model $T_{eff}$ has an opposite behavior. Note that

\section{\label{sec:Experiment}Experimental characterization of an actual barrier}
\begin{figure}
\includegraphics[width=7cm]{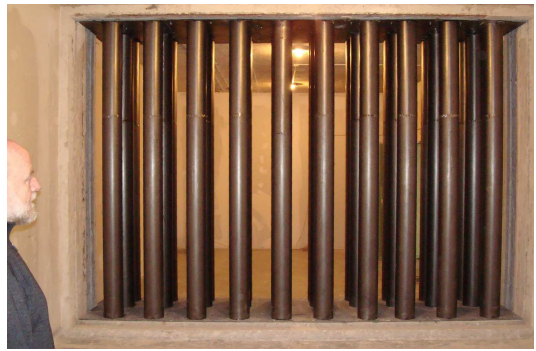}
\caption{\label{foto_camara} (Color online) Photograph taken inside the transmission chamber. The barrier is schematically described in Fig. \ref{fig:EsquemaBarrera} and has parameters: $r_1=r_2=R_3=$10cm, $r_{i1}=R_{i2}=0cm$, $r_{i3}=$4.5cm, $d_1=$30cm, $d_2=$50cm and $D=$42cm.}
\end{figure}
To validate our theoretical predictions, a 3m high by 5m long barrier has been constructed and experimentally characterized in a transmission chamber (see Fig. \ref{foto_camara}). 
Due to practical limitations, the optimized dimensions in Table \ref{tab:ResultadosOptim} have been slightly modified, so that building parameters are $R_1=R_2=R_3=10cm$, $R_{i1}=R_{i2}=0$, $R_{i3}=4.5cm$, $d_1=30cm$, $d_2=50cm$ and $D=42cm$. 
Theoretically the expected $DL_R$ of this barrier differs by less than 0.5 dB with respect to the optimum value reported in Table \ref{tab:ResultadosOptim}.

Coefficients $R_{i}$ have been measured by following the ISO140-3 norm \cite{ISO140}.
In brief, a loudspeaker is placed at the source room of the transmission chamber and generates a white noise that becomes a diffuse sound field due to the multiple reflections on the room walls. 
Sound levels in source and receiver rooms ($L_1$ and $L_2$, respectively) are acquired with moving microphones in such a manner that the sound field is averaged in time and space. Coefficients $R_i$ for each normalized third octave band are obtained as
\begin{eqnarray}
\label{eq:R_i}
R(dB)=L_1-L_2+10log\left ( \frac{S}{A} \right ),
\end{eqnarray}
where $S$ is the surface of the sample and $A$ is the equivalent absorption area of the receiver room. The parameter $A$ is obtained through the reverberation time $T$ measured in the receiver room applied to Sabine's equation
\begin{eqnarray}
A=\frac{0.16V}{T}
\end{eqnarray}
being $V$ the volume of the receiver room.
 
\begin{figure}
\includegraphics[width=7cm]{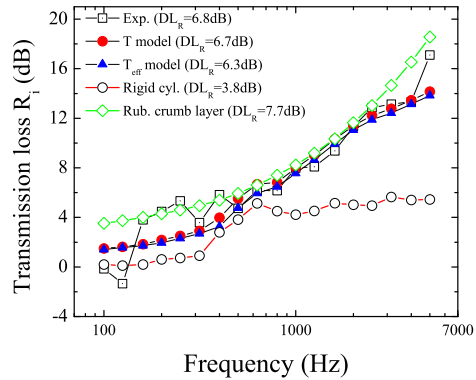}[t]
\caption{ (Color online) $R_i$ coefficients and quality factors (measured and simulated) of barrier characterized in the transmission chamber. Coefficients for a barrier made of only rigid cylinders with the same external radii and for a flat barrier with the same quantity of rubber crumb per unit length are also depicted for comparison.}
\label{fig:Ri}
\end{figure}

Figure \ref{fig:Ri} shows the $R_i$ coefficients measured and calculated for models $T$ (bold circles) and $T_{eff}$ (bold triangles). 
According to the absorptive properties of rubber crumb, the responses of attenuation obtained increase as frequency does \cite{San10}. As previously discussed, this barrier employs absorption as the main attenuation phenomenon while reflectance plays a minor role. Note that although a bandgap can be found in the reflectance spectrum, it does not appear if Fig. \ref{fig:Ri} because the angular and frequency average performed on the $R_i$ coefficients smoothes the response.
The agreement between theoretical predictions and experiment (hollow squares) is remarkable, being the obtained experimental value of $DL_R=6.78dB$, close to those predicted by the two models employed. 
Theory and experiment differ only at low frequencies where normal modes of the room become important, resulting in a not completely diffuse acoustic field. 
Moreover, as discussed above, both models produce similar curves due to the reduced reflectance of the barrier (see Eq. \eqref{Teff}).

For comparison purpose, the case of a sonic crystal barrier made of only rigid cylinders (with the same external radii) has been also considered and its coefficients are also depicted in Fig. \ref{fig:Ri}. 
Note that its corresponding quality factor has been strongly reduced in comparison with the rubber crumb barrier, where absorption is the mechanism leading to the broadband sound attenuation needed for traffic noise control.
 
Figure \ref{fig:Ri} also shows the coefficients calculated for a flat panel of rubber crumb with thickness $d=20.9cm$, thus having the same amount of this material per unit length than the measured barrier. 
These coefficients have been also obtained by using Eq. \eqref{Tav}, in which the transmission $T_{SC}$ is replaced by that calculated for a slab made of a dissipative material (the rubber crumb) \cite{Bre80} 
\begin{eqnarray}
\label{eq:T_layer}
T_{RC}(\theta)=\left | e^{-\gamma d \cos(\theta)}\frac{1-R^2}{1-R^2e^{-2\gamma d \cos(\theta)}} \right |^2,
\end{eqnarray}
where $\gamma=\alpha+jk$ is the exponential propagation in rubber crumb including attenuation effects and $R$ the reflectance between two semi-infinite layers of air and rubber crumb. 
Note that the $DL_R$ value of the flat barrier is slightly higher than that for the sonic crystal barrier, but this small difference is clearly compensated with the improvement of flow resistance as well as aesthetic aspects of the sonic crystal barrier. 

\section{\label{sec:Conclusion}Summary and conclusions}
In summary, we have used an optimization procedure to obtain optimal traffic noise barrier designs based on sonic crytals that attenuate efficiently broadband noise. 
The procedure uses the quality index $DL_R$ given by the European normative . 
Since standardized measures assumes an acoustic diffuse field, a theoretical framework to predict the behavior of sonic crystals made of rubber crumb in this kind of field is proposed. 
In order to consider realistic situations, multiple reflections between parallel barriers have also been taken into account in the calculations, solving three canonical models of transmission. 
It is shown that, in general, the optimization algorithm tends to make the barrier as dense as possible, although large differences can be found between results from each model of transmission. 
As a consequence, we conclude that schemes based on multiple reflections always prefer barriers mainly absorptive. 
On the contrary, if multiple reflections are neglected, reflectance becomes a useful attenuation mechanism so reflective barriers are obtained. 
The experimental characterization of a prototype barrier constructed in a transmission room has given a strong support to the simulation algorithm here reported. 
Finally, our optimization algorithm can be extended in designing sonic crystal-based barriers to attenuate other types of broadband noise, as those coming from industrial machinery, with well defined (stationary) profiles. 

\begin{acknowledgments}
Work supported by the Spanish MICINN under contracts TEC2010-19751 and CSD2008-66 (CONSOLIDER program), and by the U.S. Office of Naval Research under Grant  N000140910554. We acknowledge F. Simon for data acquisition and D. Torrent for useful discussions.  
\end{acknowledgments}

\bibliography{bibliografia}
\bibliographystyle{unsrt}
\end{document}